# Cybernetics and the Future of Work

Ashitha Ganapathy
*School of Cybernetics*
The Australian National University
Canberra, Australia
Ashitha.Ganapathy@anu.edu.au

Michael Timothy Bennett
*School of Computing*
The Australian National University
Canberra, Australia
Michael.Bennett@anu.edu.au

*Abstract*— The disruption caused by the pandemic has called into question industrial norms and created an opportunity to reimagine the future of work. We discuss how this period of opportunity may be leveraged to bring about a future in which the workforce thrives rather than survives. Any coherent plan of such breadth must address the interaction of multiple technological, social, economic, and environmental systems. A shared language that facilitates communication across disciplinary boundaries can bring together stakeholders and facilitate a considered response. The origin story of cybernetics and the ideas posed therein serve to illustrate how we may better understand present complex challenges, to create a future of work that places human values at its core.

## I. Introduction

The 'Future of Work', a phrase frequently used by academia, industry and futurists, is an umbrella term that encompasses globalization, digitalization, upskilling, employment protection, social security [1] and more within its scope. Subject to a great deal of prognostication pre-covid, the recent disruption to work norms has given organizations, employees, and futurists new insights and an opportunity to reflect on their predictions, to envision new futures.

## II. Future of Work

Systems of many interconnected components with non-linear relations are often difficult to recognize, manage and predict [2]. The future of work is such a system, composed of interconnected subsystems, each with its own feedback mechanisms. A three-dimensional approach of work, the workforce, and the workplace [3] is often used to simplify and understand the problem. However in this paper, the boundaries and scope have been expanded to view the future of work as consisting of four subsystems; social, technological, economic, and environmental, each of which have their own layers. In fig. 1, the relationships between each of the subsystems and layers have intentionally not been mapped to reduce visual complexity, but it is imperative to understand that a decision taken in one subsystem may have consequences within the three other subsystems, in the manner of a fully connected graph.

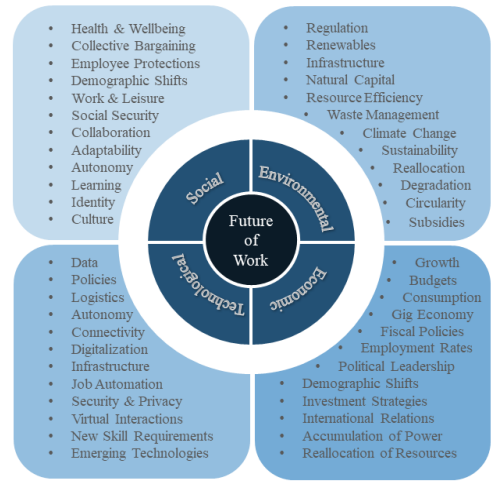

Fig. 1. Simplified diagrammatic representation of the future of work

## III. Cybernetics and its Origin

The term cybernetics was used by Norbert Wiener to name a new discipline that touched upon multiple other disciplines, that flourished on the no man's land between established fields. It required not only expertise in one's own field, but sufficient knowledge of other disciplines to understand and analyse them. The foundation of cybernetics is often attributed to the Macy conferences (1946-53) [4]. These post World War II conferences brought together specialists from disparate fields to develop a shared language that allowed for meaningful conversations; between experts, disciplines and across social boundaries.

### A. Diverse Perspective

Macy conferences engaged mathematicians, psychologists, biologists, physiologists, physicists, anthropologists, engineers, and linguists. While majority of the participants were associated with American institutions, there were also attendees from institutions in the United Kingdom, Mexico, Switzerland, Holland, and Israel [5]. Such diversity of expertise allowed for the exploration and application of concepts from a wide variety of disciplines. From living beings and machines, to economic and cognitive processes, these early studies influenced thinking in biology, neurology, sociology, ecology, economics, politics, psychoanalysis, linguistics and computer science [4,5].

### B. Shared Language

To converse across disciplinary boundaries, quickly develop an understanding of unfamiliar concepts and to add value to the debate, a shared language was developed in the conferences. Cybernetics focuses on behaviours. It does not ask "what is this thing?" but "what does it do?" [6, p.1]. Behaviour was described in terms of feedback, information and messaging. Defining concepts in terms of behavioural patterns meant they were more relatable and applicable not just to machines but to human, animal, and social phenomena [5].

*C. The Purpose*

Wiener's article in the Atlantic 'Scientists Rebel' urged scientists to take responsibility for the consequences of sharing their ideas. He argued the provision of scientific results and ideas was not an innocent act, that it was essential to know how research may be used and by whom. When Wiener was asked to share data on the target-seeking missile project he had worked on, he refused. He believed his research would result in civilian deaths, an act in which he wanted no role [7]. While others were enchanted by new technologies and their possibilities [8], Wiener stressed the importance of "know-what" alongside "know-how", a reflection on the purpose itself rather than being driven only by how to achieve it [9, pp. 184-185]. Wiener understood that seemingly innocuous scientific progress may have unforeseen consequences, both great and terrible. While technology is neither good nor evil the society in which it is embedded is complex and may employ well intended research for undesirable ends. He understood that technology could fundamentally change the nature of work. He made efforts to engage with labor unions but embroiled in more immediate concerns they were unprepared to consider political, technical, and sociological issues which questioned the very existence of labor.

Wiener also stressed the importance of developing more comprehensive methods with which to evaluate research. Assessment solely from an economic perspective could ignore potentially dire consequences for society. Instead, he believed that human values should be at the core of technological assessment and discussion [9].

## IV. Application of cybernetics to reimagine the future of work

The future of work has a long history. In the 18th century, the steam engine replaced man as a source of power [9]. In 1911, Taylor commodified labor, and glorified productivity [10], in 1984, Noble warned that a primitive enchantment with automation coupled with capitalism could have disastrous consequences for labor and society [8]. In 2005, Friedman described the workforce as a part of complex globalized supply chains, more interconnected than ever before, feeling either threatened or optimistic [11]. More recent pre-covid literature ranged from warnings of a dystopian future [12], to more practical considerations of upskilling to match the ever-changing labor market [1]. In only a year the pandemic fundamentally altered the nature of this prognostication. Integral aspects of work, the very definition of a workplace and a workforce are now being questioned. Organizations forced into rapid, unplanned transitions to remote work are now designing what their new normal will look like. Reimagining and redesigning the future of work is complex. To design such systems, it helps to be able to draw upon every tool available, to consider a variety of perspectives. However, considering many and varied perspectives together can be difficult, for to communicate different stakeholder groups must speak the same language. The Macy conferences illustrate how this may be done. By adopting and further developing their ideas and shared language, polarized positions may be compared side by side to identify the right questions, debate, and reimagine.

*A. Who Designs It and for Whom?*

Designing a future of work requires stakeholders to work together, yet those stakeholders may be at cross purposes. Governments, organizations, think tanks, policy makers, economists, social scientists, environmentalists, citizens, and the workforce may have conflicting goals. While meeting every goal may be impossible, assessing tradeoffs and understanding the impact of each of those tradeoffs in relation to different time frames and amongst different stakeholders will be useful in guiding decisions for the future.

Parties directly impacted by any design should be actively sought out and engaged in the design process. Post-covid, organizations have resorted to collecting data from employees via interviews and surveys to gauge the needs and desires of their workforce. Such data collection is not without risk. Not all organizations are prepared to collect, store, and analyze data of a sensitive nature. Psychological, personal and household related information. In the event of a security breach, the possession of such data poses a risk to individual employees and a liability for the organization.

If a choice must be made between mutually exclusive goals held by opposing stakeholders, we must consider who will be making that choice. Wiener openly expressed his opposition to the accumulation of power in the hands of a few. He questioned decisions that place personal gain above any duty to society at large, and subsequently exhorted collective movements ensuring the integrity of institutions and working conditions [9]. Among other factors, $20^{th}$ century automation weakened labor unions [8]. Collective bargaining is an important tool. It has helped shape rights, regulate the use of new technologies, or foster labor market security and adaptability [1]. The future of work is not a linear process with a destination but more an iterative process, a loop in which each point in time presents us with a new starting point to consider. Any reimagined future must account for how the environment is already changing, and somehow integrate a mechanism for collective bargaining within a globalized economy that is already seeing an increase in freelance, gig, and other non-standardized work. Collective bargaining becomes more complex when workers are only connected indirectly within large supply chains, whilst being disconnected geographically, culturally, and socially.

*B. Systems*

Wiener understood the role of technology as a catalyst for political, economic, and social change. Instead of considering each part only in isolation, we should also examine the consequences of any decision for the larger system [13]. Identifying levels of change and the associated emergent properties of the system can help avoid long-term unintended consequences [14,15]. Wiener pointed out that mining and lumber industrialists had often failed to properly consider consequences of decisions based upon financial considerations alone [9].

The debate on remote work illustrates several aspects of interconnectedness, the most obvious being the absence of a dedicated working environment separate from home, of a social life centered on the office; a disruption to place attachment. Such "disruption to continuity, acts to close a category of past experiences (interactional past), but it also acts to alter one's expectations for the future (the interactional potential of a site)" [16, pp. 8-9]. An acknowledgement that employees have not only lost a part of their past but are anxious about the uncertainty that lies ahead, will require organizations to empathize and provide new types of support to their workforce. Other consequences could include large-scale demographic shifts, a demand for new technologies to enhance collaboration and improve connectivity, and the legitimization of new intrusive surveillance tools to manage a distributed workforce. Adoption of snooping software to detect laziness [17], to determine productivity based on body language [18] and to monitor employees through webcams carry high risks [19] and are intrusive in nature. Regulatory bodies will be required to draw a line between overly intrusive and justifiable collection of information. The environmental implications of remote work vary geographically based on patterns of energy consumption, major energy sources and weather patterns [20].

*C. Values at the Core*

Values are what an individual or group of individuals consider important in life. Technology is not value neutral, but an expression of the culture and values of those who designed it [21].

When we design the future of work, stakeholders should be aware of their values, and how those values may or may not be aligned with the purpose of the system in which they're engaged. Following are four examples of values which may influence design choices. These values are not an attempt at making an all-inclusive list and we acknowledge that the process of value extrication should include both direct and indirect stakeholders [21].

*1) Trust* – Research has highlighted the importance of trust within groups and organizations as an essential ingredient for teams to feel psycologically safe. Trust is defined as the expectation that others' future actions will be favorable to one's interests, such that one is willing to be vulnerable to those actions [22]. The level of trust in turn impacts the level of autonomy within organizational structures. Research has shown that autonomy at work is known to improve worker performance and well-being [23]. It defines the level of control one has over their work. It can vary significantly in different roles based on the nature of work. For a freelancer, it is the option of choosing when to work and what to work on and for a full- time employee it maybe choosing how to work on the tasks and when to work on it (within the constraints of larger team deliverables). Elements of the factory system where workers are viewed as inherently lazy [10] and need constant monitoring are ingrained in practices, processes, and our culture. The development and suggestion of snooping software is an example of the need to control the workforce even at the cost of their work performance and well-being.

In our new future, can our default position be that of trust?

*2) Dignity* - Novogratz views dignity, not wealth, as the inverse of poverty. Human dignity depends upon the ability to make decisions, to self direct and affect changes [24]. Wiener perceived the automation of labor as economically akin to legalising slave labor [9]. To compete with slave labor, a worker may be forced to accept poor working conditions. Little compensation, longer hours, no time to invest in family or even building a better skill set may become the norm. In other words, just surviving. However, the pandemic has provided an opportunity to reflect on the working habits of the past and reimagine them. Placing dignity at the forefront, the conversation can shift from surviving to thriving [3]. From competing with machines to being human and working alongside machines, to celebrating humanness and the importance of work to identity, to creativity and to society.

In our new future, can dignity of work be afforded to all?

*3) Empathy* – In a world as polarized as ours, there is a tendency to form and join homogenous groups with whom we share obvious traits such as education, religion, politics and wealth. Such fragmentation may be exacerbated by globalization, digitization or other percieved threats to both culture and identity. Demagogues encourage such polarization, amplified by the algorithms driving social media news feeds maximizing engagement through anger inducing headlines and misinformation, we live in an age where empathy can be hard to come by. Empathy is a combination of emotional and cognitive processes. Using a combination of definitions from philosophy, religion and psychology it is defined as "the ability to imagine or feel ourselves in the position of others" [25, p.57]. It plays an important role in society, "enabling sharing of experiences, needs, and desires between individuals and providing an emotional bridge that promotes prosocial behavior". Research shows empathy is mutable and can be developed [26, p.74]. However, when people are overwhelmed and experiencing burnout, their capacity to empathise with others declines. As per the World Economic Forum, 88% of US workers said they were moderately to extremely stressed during the pandemic and over two-thirds said it was the most stressful time in their careers [27, p. 15]. Self-empathy, showing compassion to oneself, is essential in cultivating the ability to empathise with others. Organizations must ensure employees are supported in sustaining a capacity for empathy [26]. Direct interaction between humans has diminished in favor of human-computer interaction. We are constantly engaged with inanimate objects that serve as a barrier to more fulfilling direct socialisation. A directed effort to cultivate empathy in different levels of the workforce can ensure a more resilient and adaptable workforce.

In our new futures, how might empathy be cultivated?

*4) Sustainability* – Sustainability in this context is defined not just in environmental but economic and social terms. Environmental sustainability refers to the impact on the environment now and the future. Economic sustainability refers to the economic viability of decisions. Economic and environmental decisions cannot be viewed independent from each other, the economy cannot expand beyond the confines of ecological limits. Social sustainability refers to the health and well-being of the worker, their families and communities now and in the future. Those in poverty may be more vulnerable to the effects of environmental degradation, therefore having sustainability as a value indirectly reduces inequality. We are currently using 1.7 times the resources our planet can sustainably generate. Our current economic assessments do not effectively account for the efficient allocation of resources and the preservation of natural capital. In 2014, 40% of the global workforce was employed in industries heavily dependent on natural processes [28]. As illustrated earlier Wiener disapproved of the financially oriented assessment methods for new technoligies, and industrialists concerned only with profit maximization [9]. While the preceding values are human-centric, sustainability suggests decisions consider the environment and everything in it.

In our new future, how might sustainable practices be encouraged?

## V. Conclusion

The term 'global village', coined by McLuhan, has a rather fluid meaning. Its interpretation has and continues to change as the context within which it exists changes. He defined the concept to be the creation of shared environments of communities on a global scale through media technologies [29]. The very meaning of shared environments and media technologies in today's world have changed. The borders of a child's village today are far more expansive than those of a child born 20 years ago. However, the motivating principles of this idea remain relevant. Both Wiener and McLuhan highlighted the importance of long-term planning, assessing the consequences of present decisions for the future [9,29].